# Pseudo-Centroid Clustering

Fred Glover

Department of Electrical, Computer and Energy Engineering
School of Engineering & Science
University of Colorado
Boulder, Colorado 80309, USA
glover@colorado.edu

October 2016

## Abstract

Pseudo-centroid clustering replaces the traditional concept of a centroid expressed as a center of gravity with the notion of a *pseudo-centroid* (or a *coordinate free centroid*) which has the advantage of applying to clustering problems where points do not have numerical coordinates (or categorical coordinates that are translated into numerical form). Such problems, for which classical centroids do not exist, are particularly important in social sciences, marketing, psychology and economics, where distances are not computed from vector coordinates but rather are expressed in terms of characteristics such as affinity relationships, psychological preferences, advertising responses, polling data and market interactions, where distances, broadly conceived, measure the similarity (or dissimilarity) of characteristics, functions or structures.

We formulate a K-PC algorithm analogous to a K-Means algorithm and focus on two key types of pseudo-centroids, *MinMax-centroids* and (weighted) *MinSum-centroids*, and describe how they, respectively, give rise to a K-MinMax algorithm and a K-MinSum algorithm which are analogous to a K-Means algorithm. The K-PC algorithms are able to take advantage of problem structure to identify special *diversity-based* and *intensity-based* starting methods to generate initial pseudo-centroids and associated clusters, accompanied by theorems for the intensity-based methods that establish their ability to obtain best clusters of a selected size from the points available at each stage of construction. We also introduce a regret-threshold PC algorithm that modifies the K-PC algorithm together with an associated diversification method and a new criterion for evaluating the quality of a collection of clusters.

# 1. Introduction.

Consider a set X of n points, given by X = {$x^r$, r ∈ N = {1, …, n}}. We are interested in the clustering problem that consists of identifying a partition of X into subsets C = C(h), h ∈ K = {1,…,k} for specified values of k.

In this paper, we are specifically interested in clustering problems where the goal is to organize clusters by a criterion based on the distances of points in each cluster from a common (suitably defined) "central point," but where the points lack coordinates that make it possible to draw on the notion of a classical centroid as used in methods such as the K-Means algorithm. We emphasize that we interpret "distance" in a broad sense to encompass any value assigned to a pair of points that expresses their proximity or similarity. We may think of distance as a type of repulsion, for example, where larger distances reflect a reduced desirability of placing the two associated points in a common cluster. We also allow consideration of negative distances, which may be viewed as a form of attraction.

To handle situations where distances are not restricted to constitute a metric based on spatial coordinates of a vector, we introduce the notion of a *pseudo-centroid* (PC) to replace that of a classical centroid. This enables us to describe K-PC algorithms by analogy with the K-Means algorithm for applications in which points $x^r$ to be clustered include non-numerical coordinates (or lack coordinates altogether). Such problems occur prominently, for example, in the social media, marketing and economics areas, where distances, broadly conceived, may represent affinity relationships, psychological preferences, advertising responses, polling data, market interactions and so forth.

We establish the background and general ideas underlying pseudo-centroid clustering, and focus particularly on two forms of this clustering we call K-MinMax and K-MinSum clustering. The MinMax and MinSum-centroids make it possible to generate starting clusters for the K-MinMax and K-MinSum clustering algorithms that embody a meaningful structure in relation to the goal of producing a best collection of final clusters and which apply as well to other definitions of pseudo-centroids discussed in Section 9. We demonstrate this by providing advanced starting procedures for these clustering algorithms of two types, called diversity-based and intensity-based methods, which also offer ways to aid in determining good values of k and provide variants that can be used with K-Means algorithms. We also including special adaptive versions of the intensity-based methods designed to reduce the number of iterations subsequently performed by K-PC algorithms and to increase the likelihood of yielding final clusters that are the best achievable. The intensity-based methods are accompanied by theorems that identify the quality of clusters they produce.

# 2. Background and related research

A variety of papers have undertaken to develop measures to replace Euclidean distances, or in some cases to transform non-numerical coordinates (such as those for categorical variables) into numerical coordinates to permit Euclidean distances to be calculated. For instance, Ralambondrainy (1995) converts multiple category attributes into binary attributes (1 if a category is present, 0 if absent) to treat binary attributes as numeric in the K-Means algorithm (MacQueen, 1967). This approach needs to handle a large number of binary variables if used in data mining, where categorical attributes often involve hundreds or thousands of categories. Huang (1998) goes farther by developing ways to handle categorical variables utilizing a simple dissimilarity measure for categorical objects proposed by (Kaufman and Rousseeuw, 1990) and applies a variant of the K-Means algorithm using modes instead of means for clusters, together with a frequency-based method to update modes to minimize the clustering cost function. The foregoing approaches continue to rely on coordinates, however, rather than directly making use of distance measures to characterize a notion of centrality without the need to refer to coordinates.

Even in situations where Euclidean distances exist naturally and provide the possibility for creating centroids, a reliance on such centroids is not always desirable. For example, as illustrated in Cao and Glover (2010), centroids can create difficulties for problems where communications or travel between points must take place within a specified network, since centroids may lie in unreachable areas such as rivers or fields that cannot be traversed, or may compel some points that lie in a centroid-based cluster to cross such areas to reach other cluster points, thus linking points that are not desirable to be grouped within a cluster together.

On the other hand, there are many methods that rely on distances without making use of centroids. Such methods avoid the problems that typically arise by centroid-based methods and have a venerable tradition, although at the cost of abandoning an effort to group points with reference to a point that satisfies a meaningful alternative definition of centrality. For example, Fan (2009) describes a distance measure that modifies Euclidean distances to incorporate an "obstructed distance" between two locations if their straight line link is intersected by geographic obstacles such as rivers, mountains or highways. He then undertakes to use these distances by a method that uses a simulated annealing approach to correct distortions introduced by an attempt to use a K-Means algorithm. But no measure is offered to replace a centroid-based measure as relied on by the K-Means procedure.

In a related setting, geographic information system (GIS) technology is often utilized due to its ability to supply vital information such as geographic feature data, street network information, speed limits on street segment and lengths of street segments. This technology is also able to keep track of restrictions such as vehicle heights, weights and volumes that need to be considered by optimization procedures. For example, Estivill-Castro and Lee (2001) combine data mining and GIS as a means to consider geographic obstacles such as hills or rivers. The authors devise a clustering algorithm utilizing a Voronoi diagram to set up a topological structure for a set of points as a basis for retrieving spatial information related to various definitions of neighbors. Once again, having established the topological structure, no way of representing centrality is provided. Another use of a Voronoi diagram to yield a more appropriate space is given in Kwon et al. (2007) as a basis for a tabu search method for solving capacitated vehicle routing problems. However, relying on the Voronoi topology by itself without a new concept of central point was not sufficient to allow the approach to beat the existing benchmark results.

A different type of application is addressed by Strehl and Ghosh (2002), who develop a clustering algorithm for data mining problems found in the retail industry and some Web applications, where data reside in a very high-dimensional space. This approach introduces a similarity relationship defined on each pair of data samples and transforms the problem into one over the similarity domain so that the original high-dimensional space is no longer needed. The goal is to cluster data samples into $k$ groups so that data samples for different clusters have similar characteristics. Lacking a meaningful centrality measure, the authors formulated the clustering problem as an NP-hard vertex-weighted graph partitioning problem and developed an algorithm called Optimal Partitioning of Space Similarities Using Metis (OPOSSUM) using the Metis approach proposed by Karypis and Kumar (1998) as the multi-objective graph partitioning engine. Another recourse to a more complex model representation to cluster objects in the absence of an ability to take advantage of central points occurs in Kochenberger et al. (2005), who use clique partitioning to cluster microarray data.

A type of approach that is related to one of the types of pseudo-centroid methods described in this paper is the K-Medoid algorithm embodied in the Partitioning Around Medoids (PAM) algorithm of Kaufman and Rousseeuw (1990), and in its extended variant Clustering Large Applications CLARA). Ng and Han (2002) further extend this approach with a method called CLARANS (Clustering Large Applications based upon RANdomized Search). Specifically, our definition of a MinMax-centroid corresponds to one of the definitions of a medoid, but the medoid definition incorporated into the foregoing K-Medoid algorithms has a different foundation that stems from a coordinate-centric perspective (seeking to minimize the maximum deviation of cluster points from a classical centroid). Similarly, another treatment of medoids in Park and Jun (2009) employs a definition corresponding to our definition of a MinSum pseudo-centroid, but again resorts to a coordinate-centric perspective for its implementation and lacks the generality that makes the pseudo-centroid framework applicable to other definitions as well. Because of these differences, the K-Medoid algorithms cited above operate differently from the K-MinMax and K-MinSum algorithms and fail to gain access to associated starting algorithms and their associated theorems described in subsequent sections.

It is important to point out that not all forms of clustering can benefit from exploiting a measure of centrality. A reliance on centrality is purposely missing in the case of the spanning tree methods which provide an ability to generate clusters that may be embedded in others (see, e.g. Xu et al., 2001; Paivinen, 2005; Shamsul, et al., 2011). Similarly, centrality measures are not relevant to a variety of applications approached by the use of hierarchical clustering methods (Anderberg, 1973, Jain and Dubes, 1988) or by the use of methods such as the cohesive clustering approach of Cao et al. (2015). Many other examples of these types exist. At the same, centrality plays a vital role in numerous clustering problems and hence constitutes the focus of this paper.

The remainder of this paper is organized as follows. Section 3 introduces the basic concepts underlying pseudo-centroids (PCs) and their embodiment in a K-PC algorithm. Sections 4 and 5, respectively, introduce the two instances of a PC-centroid called the MinMax-centroid and the MinSum-centroid, together with their associated K-MinMax and K-MinSum algorithms. Section 6 addresses the topic of advanced starting methods to generate initial sets of points for the K-MinMax and K-MinSum algorithms, and describes the class of diversity-based starting methods that also can help to determine good values of k for the number of clusters to be generated. Section 7 then introduces the class of intensity-based starting methods with a focus on the primary intensity-based methods. These methods have an ability to successively generate clusters that are locally optimal for the chosen number of elements to be included in a cluster, given the set of elements available at the current stage of construction. The more advanced adaptive intensity-based starting methods are introduced in Section 8, which allow the number of elements in a cluster to be generated adaptively at each stage. Finally, a strategic variant of the K-PC algorithm, called the regret-threshold PC algorithm, is introduced in Section 9, together with a diversification method and a new criterion for evaluating a collection of clusters. The paper concludes with observations concerning other types of pseudo-centroids and applications involving metaheuristic optimization in Section 10.

## 3. Pseudo-Centroids and a K-PC Algorithm

Let $d(i,j)$ denote a distance measure defined on the elements $x^i, x^j \in X$. We follow the convention $d(i,i) = 0$, but do not require that $d(i,j)$ satisfies the conditions to be a metric or that it be non-negative. For convenience, as in the case of $d(i,j)$, we allow points $x^i$ and $x^j$ to be referenced by their indexes. Let C denote an arbitrary set of points (indexes) in N. In the following we refer to a *separation measure* Separate(i: C) that represents the separation of i from other points $j \in C \setminus \{i\}$. As developed here, Separate(i: C) is a function of the distances $d(i,j)$ for $j \in C \setminus \{i\}$ and specific instances of Separate(i: C) are given in the following sections.

*Definition PC*: A *Pseudo-Centroid* of C is a point $i^* \in C$ that gives the smallest value of Separate(i: C); i.e.:

$$\text{Separate}(i^*, C) = \text{Min}(\text{Separate}(i: C): i \in C)$$

The value Separate($i^*$: C) will be called the *span* of C, and denoted Span(C). The span value may be thought of as a "radius" about $i^*$ (defined by reference to the set separation measure Separate(i: C)), conceiving $i^*$ as a "central point," such that all points $j \in C \setminus \{i\}$ lie within the separation Span(C) from $i^*$.

An important characteristic of pseudo-centroids, as will be evident from the two primary instances we discuss later, is that unlike ordinary centroids they may not be unique. We account for this by including a variation of our algorithms based on pseudo-centroids that refers to the set C* of all points in C that qualify as a pseudo-centroid of C; i.e., C* = {i $\in$ C: Separate(i: C) = Span(C)}. The point $i^*$ in Definition PC thus belongs to C*. We first describe the K-PC algorithm by assuming we arbitrarily single out a particular element of C* to be identified as $i^*$, in the case where more than one such element exists, and then describe the variant that references all of C*.

**K-PC Algorithm**

0. Begin with a collection of k points $N(K) = \{i(h), h \in K\} \subset N$, and assign each of the n – k points $j \in N \setminus N(K)$ to the point $i = i(h)$ that is closest to it by the measure $d(i,j)$. (That is, $j \in N \setminus N(K)$ is assigned to a point $i(h) = \arg\min (d(i(h),j): h \in K)$.) Identify an initial collection of clusters $C(h), h \in K$, each consisting of $i(h)$ and the points assigned to it.

1. Identify a pseudo-centroid $i^*(h)$ for each cluster $C(h)$ in accordance with Definition PC; i.e., Separate($i^*$: C) = Span(C) for $i^* = i(h)$ and $C = C(h)$. Denote this set of pseudo-centroids by $N^*(K) = \{i^*(h): h \in K\}$.

2. Reassign each of the n – k points $j \in N \setminus N^*(K)$, to the pseudo-centroid $i = i^*(h)$ that is closest to it by the measure $d(i,j)$, thus assigning each j to an associated new cluster $C(h)$ that contains $i^*(h)$.

3. Repeat Steps 1 and 2 (creating a new set of pseudo-centroids at Step 1) until the set $N^*(K)$ does not change or until a chosen iteration limit is reached.

As the foregoing description shows, the KC-PC algorithm closely follows the format of the K-Means algorithm (considering both in their simplest forms) except for the reference to the separation measure Separate(i: C) and the introduction of pseudo-centroids in place of classical centroids.

The variant of the foregoing algorithm that incorporates the set $C^*$ of all pseudo-centroids amends the definition of $N^*(K)$ so that it becomes the union of the sets $C^*(h), h \in K$. Then Steps 1 and 2 of the K-PC algorithm become as follows.

**Modified Steps 1 and 2 to incorporate all pseudo-centroids**

1. Identify the set $C^*(h)$ of pseudo-centroids for each cluster $C = C(h)$ by $C^*(h) = \{i \in C:$ Separate(i: C) = Span(C)\}. Denote this set of pseudo-centroids by $N^*(K) = \cup(C^*(h): h \in K)$.

2. Reassign each of the $n - |N^*(K)|$ points $j \in N \setminus N^*(K)$, to the pseudo-centroid $i^* \in N^*(K)$ that is closest to it by the measure $d(i,j)$, thus assigning each j to the associated new cluster $C(h)$ such that $i^* \in C^*(h)$ (retaining each $i^* \in C^*(h)$ in the this new cluster).

An additional variant of the K-PC algorithm allows all n points $j \in N$ to be reassigned in Step 2 instead of only the points $j \in N \setminus N^*(K)$. If the distances $d(i,j)$ are non-negative, and we interpret the convention $d(i,i) = 0$ to permit $d(i,i)$ to be slightly negative, then this implies that a pseudo-centroid will be re-assigned to itself, and hence the outcome of Step 2 will not change for this variant. However, in the case where negative distances exist, the indicated variant could re-assign a pseudo-centroid to a different pseudo-centroid, and thus reduce the number of clusters generated. Later we discuss other variants of the K-PC algorithm, including a Regret-Threshold PC algorithm, which in turn suggest new variants of the K-Means algorithm.

Finally, we note that the classical form of the K-Means algorithm terminates when an objective function expressed as the sum of the distances, or squared distances, of each point from its assigned centroid, summed over all centroids, reaches a local minimum. An analogous (though clearly not equivalent) objective function for the K-PC algorithm is to minimize

$$\sum(\text{Span}(C(h)): h \in K)$$

This function may likewise be replaced by one in which the Span(C(h)) values are squared. Thus an alternative termination point for the K-PC algorithm is to terminate when this function attains a local minimum, where the next iteration causes the function to increase. This objective may also be modified to replace Span(C(h)) by Span(C(h))/|C(h)|, thus seeking to minimize the sum of the "average" span values over the clusters. In Section 9 we present a somewhat different function for evaluating the clusters which can also be used to identify good values of k.

## 4. The MinMax-Centroid and the K-MinMax Algorithm

The first principal instance of a pseudo-centroid we consider, the MinMax-centroid, arises by defining Separate(i: C) = Max(d(i,j), j ∈ C \ {i}), which we also denote by MaxDist(i,C). Then we obtain the instance of Definition PC given by

*Definition MinMax*: A *MinMax-Centroid* of C is a point i* ∈ C whose maximum distance from all other points in C is smallest; i.e.:

$$\text{MaxDist}(i^*: C) = \text{Min}(\text{MaxDist}(i: C): i \in C)$$

Then Span(C) = MaxDist(i*: C) and for this case will be denoted MMSpan(C).

The statement of the K-MinMax Algorithm then becomes essentially the same as the statement of the K-PC algorithm, upon replacing "pseudo-centroid" with "MinMax-centroid," "Separate(i: C(h))" with "MaxDist(i: C(h))" and "Span(C) with MMSpan(C)."

**K-MinMax Algorithm**

0. Begin with a collection of k points N(K), and assign each of the n – k points j ∈ N \ N(K) to the point i = i(h) ∈ N(K) that is closest to it by the measure d(i,j). Identify an initial collection of clusters C(h), h ∈ K, each consisting of i(h) and the points assigned to it.

1. Identify a MinMax-centroid i*(h) for each cluster C(h); i.e., MaxDist(i*: C) = MMSpan(C) for i* = i(h) and C = C(h). Denote the set of MinMax-centroids by N*(K) = {i*(h): h ∈ K}.

2. Reassign each of the n – k points j ∈ N \ N*(K), to the MinMax-centroid i = i*(h) that is closest to it by the measure d(i,j), ), thus assigning each j to an associated new cluster C(h) that contains i*(h).

3. Repeat Steps 1 and 2 (creating a new set of MinMax-centroids at Step 1) until the set N*(K) does not change or until a chosen iteration limit is reached.

The time complexity of the K-MinMax algorithm compared to that of the K-Means algorithm depends on the dimension d of the points i ∈ N (vectors $x^i$, i ∈ N), which are used to compute a centroid in the K-Means algorithm, as determined by the execution of Step 1 (since Steps 0 and 2 are essentially the same in both algorithms). The complexity of identifying a MinMax-centroid of a cluster C(h) in Step 1 of the K-MinMax algorithm is $O(|C(h)|^2)$, while the complexity of identifying an ordinary centroid with the K-Means algorithm is O(|C(h)|d). If the distances d(i,j) must be precomputed from coordinates, rather than being given directly from the problem data base for the application of concern, then this will inflate the the complexity of the K-MinMax algorithm. In this respect, it may be noted that even where distances rely on coordinate calculations, commercial enterprises or government agencies may prefer not to disclose the information contained in these coordinates, and therefore will preprocess the data to produce distances before passing them along to a clustering algorithm.[1]

Independently of such considerations, the computational effort of Step 1 in the K-MinMax algorithm can potentially be improved as follows:

*Accelerated update for Step 1.* For each MinMax-centroid i* = i*(h) identified in Step 1, let *match*(i*) be a point j* that determines the MMSpan(C(h)); that is, d(i*, j*) = MaxDist(i*: C(h)). If j* belongs to the new cluster C(h) produced in Step 2, then the computation to identify the MinMax-centroid of C(h) on the next execution of Step 1 can restrict attention to the set of points $C^+(h)$ that were added to C(h) in Step 2. Specifically, for each i ∈ $C^+(h)$ compute the value MaxDist(i: C(h)) = Max(d(i,j): j ∈ C(h) \ {i}) and from these values identify i'(h) = arg min(MaxDist(i: C(h)): i ∈ $C^+(h)$). Then the MinMax-centroid of the new C(h) passed from Step 2 to Step 1 is either i'(h) or i*(h), depending on which of MaxDist(i'(h): C(h)) and MMSpan(C(h)) (= d(i*(h),j*)) is smaller.

The modification of the K-MinMax algorithm to refer to the full sets of MinMax-Centroids C*(h), when these may contain more than one element, changes Steps 1 and 2 as indicated earlier. In this case the changed form of Step 1 identifies C*(h) by C*(h) = {i ∈ C: MaxDist(i: C) = MMSpan(C)} for C = C(h) (together with redefining N*(K) = ∪(C*(h): h ∈ K)) and the changed form of Step 2 reassigns each of the n – |N*(K)| points j ∈ N \ N*(K), to the MinMax-centroid i* ∈ N*(K) that is closest to it, creating new clusters C(h) (which, respectively, inherit the previous elements of C*(h)).

The Accelerated Update of Step 1 can be applied to the situation where we keep track of the sets C*(h) as a result of the fact that MMSpan(C(h)) is the same value regardless of which element i*

---

[1] Communication from B. Cao, regarding applications of the method of Cao et al. (2015) to commercial applications.

∈ C*(h) is used to identify it, and it is only necessary to retain a single associated element j* in C(h) (for some i*) to imply the value MMSpan(C(h) (previously = d(i*,j*)) remains at least as large in the new C(h) as it was before. Likewise we may refer to a single index i'(h) to compare the two distances MaxDist(i'(h): C(h)) and MMSpan(C(h)). Then the new set of MinMax-centroids will be determined according to which of these distances is smaller. Since all MinMax-centroids of the old C(h) are required to be carried forward as elements of the new C(h), the new set C*(h) will contain the old C*(h) plus any additional points of $C^+(h)$ that qualify to be included.

## 5. MinSum-Centroids and a K-MinSum Algorithm

The second type of pseudo-centroid we focus on, called a MinSum-centroid, results by defining defining Separate(i: C) = SumDist(i,C), where SumDist(i,C) = $\sum$ (d(i,j), j ∈ C \ {i}). Then we obtain the instance of Definition PC given by

*Definition MinSum*: A *MinSum-centroid* of C is a point i* ∈ C whose sum of distances from all other points in C is smallest; i.e.:

$$\text{SumDist}(i^*: C) = \text{Min}(\text{SumDist}(i: C): i \in C)$$

Then SumDist(i*: C) identifies the value of Span(C), which in this case will be denoted MSSpan(C).

A weighted version of a MinSum-centroid can be created without introducing weights separately but simply by modifying the distances. Specifically, suppose it is desired to attach a weight w(D) to each d(i,j) such that d(i,j) = D. Then it suffices to create the new value d'(i,j) := w(D)d(i,j) for D = d(i,j) and then use d'(i,j) in place of d(i,j). To keep the ordering the same for d'(i,j) and d(i,j) we would specify that w(D) is a monotonically increasing function of D. This occurs, for example, for non-negative distances by setting w(D) = $D^p$ where p ≥ 0. (E.g., setting p = 1 yields d'(i,j) = $d(i,j)^2$.)

For weights w(D) that grow fast enough (as where w(D) = $D^p$ for a very large power p), the MinSum-centroid becomes a special instance of the MinMax-centroid – one where ties among candidates for i* are implicitly broken by a lexicographic rule: if the largest d(i,j) values are the same, then the second largest d(i,j) values are compared for the tied elements, and if ties remain at this level, the third largest d(i,j) values are compared for the resulting tied elements, and so on. However, a scheme employing such large weights is very likely to produce d(i,j) values too large to be easily handled computationally, thus making the direct MinMax-centroid preferable to the weighted MinSum-centroid for problems where maximum distances are more relevant than (weighted) sums of distances.

On the other hand, when weights of reasonable size are employed, the structures of clusters produced by a K-MinMax Algorithm and a K-MinSum Algorithm can differ appreciably, and hence it becomes of interest to compare the two approaches to determine which may have advantages in particular contexts.

While the statement of the K-MinSum Algorithm can be readily inferred from the statement of the K-PC Algorithm and our preceding definitions, we give explicit instructions for the K-MinSum Algorithm as follows. By this means, we are able to make observations about the K-MinSum Algorithm that differ from those concerning the K-MinMax Algorithm, particularly in the context of advanced starting methods of later sections.

**K-MinSum Algorithm**

0. Begin with a collection of k points N(K), and assign each of the n − k points j ∈ N \ N(K) to the point i = i(h) ∈ N(K) that is closest to it by the measure d(i,j). Identify an initial collection of clusters C(h), h ∈ K, each consisting of i(h) and the points assigned to it.

1. Identify a MinSum-centroid i*(h) for each cluster C(h); i.e., SumDist(i*: C) = MSSpan(C) for i* = i(h) and C = C(h). (SumDist(i: C(h)): i ∈ C(h)). Denote the set of MinSum-centroids by N*(K) = {i*(h): h ∈ K}.

2. Reassign each of the n − k points j ∈ N \ N*(K), to the MinSum-centroid i = i*(h) that is closest to it by the measure d(i,j), ), thus assigning each j to an associated new cluster C(h) that contains i*(h).

3. Repeat Steps 1 and 2 (creating a new set of MinSum-centroids at Step 1) until the set N*(K) does not change or until a chosen iteration limit is reached.

The time complexity of the K-MinSum algorithm is the same as that of the K-MinMax algorithm, and hence the comparison to the complexity of the K-Means algorithm again depends on the dimension d of the points i ∈ N (vectors $x^i$, i ∈ N) involved in computing a centroid, as determined by the execution of Step 1. Specifically, the different complexities for executing this step are $O(|C(h)|^2)$ for the K-MinSum algorithm and $O(|C(h)|d)$ for the K-Means algorithm.

The computational effort of Step 1 in the K-MinSum algorithm can potentially be improved by a somewhat different method than in the case of the K-MinMax algorithm.

*Accelerated update for Step 1 of the MinSum Algorithm.* Let C' = C'(h) denote the set of points C(h) before executing Step 2 and C" = C"(h) denote the set of points after executing Step 2. Let $C^o$ = C' ∩ C" (the points that remain in C(h) after the reassignment of Step 2), $C^+$ = C" \ C' (the points added to C(h) in Step 2) and $C^-$ = C' \ C" (the points dropped from C(h) in Step 2). Assume the values SumDist(i,C') have been saved before executing Step 2. Then for i ∈ $C^o$, the new SumDist(i: C(h)) value is given by SumDist(i: C") = SumDist(i,C') + ∑(d(i,j): j ∈ $C^+$) − ∑(d(i,j): j ∈ $C^-$). This update can be useful for saving computation on iterations where some of the clusters C(h) do not greatly change their composition by Step 2 (i.e., where $C^o$ is relatively large compared to $C^+$ and $C^-$).

We now turn to one of the main features of the K-PC algorithm that motivates our study, which is the ability to derive an initial collection of clusters and their pseudo-centroids by means of special types of advanced starting algorithms.

## 6. Advanced Starting Methods

We introduce two types of advanced starting methods, called Diversity-Based Starting Methods and Intensity-Based Starting Methods,[2] to generate an initial set of points $N(K) = \{i(h), h \in K\}$ for the K-MinMax and the K-MinSum algorithms. The diversity-based methods are simpler and require less computation than the intensity-based methods, and apply equally to the MinMax and MinSum cases. (These methods can also be used to generate starting points for the K-Means algorithm.)

The more advanced intensity-based methods take slightly different forms for the MinMax and MinSum cases. Their greater computational complexity makes them less suited for large clustering problems, though they afford an opportunity to generate a better set of initial points to become $N(k)$ in the K-MinMax and K-MinSum algorithms. They also generate MinMax and MinSum-centroids to compose the set $N^*(k)$ for the first execution of Step 2, and simultaneously produce associated clusters $C(h)$.

The intensity-based methods are of two types, which we call Primary and Adaptive. Both generate a best cluster of a given size from the elements available at each step. This does not assure the final collection of clusters will be optimal since the cluster size may not be chosen correctly and a best cluster at a given stage may not be a globally best choice over all stages. Nevertheless, it increases the likelihood that the quality of the initial clusters will lead to reducing the overall number of iterations of the K-PC algorithm and improve the quality of the final clusters produced.

The remainder of this section is devoted to the diversity-based starting methods. The intensity-based starting methods are then described in the two subsequent sections, introducing the Primary Intensity-Based Algorithms in Section 7 and the Adaptive Intensity-Based Algorithms in Section 8.

### Diversity-Based Starting Methods

The diversity-based starting methods constitute an elaboration of an approach suggested in Glover (1994) which may be formulated in the present context by viewing N as the set from which to create a subset N(K) of k elements. In this setting, the approach may be described as starting with an arbitrary seed element from N as the first element of N(K) and then sequentially selecting each of the remaining k – 1 elements to maximize its minimum distance from the elements chosen so far. We first depict a simple instance of this method based on the following terminology. Let $H_o = \{i(h): h = 1, \ldots, k_o\}$ denote the set of starting points selected as of iteration $k_o$, where $k_o$ ranges from 1 to k. As the points i(h) are successively selected from N and added to $H_o$, we also refer to the set $I_o = N \setminus H_o$ which identifies the points remaining in N as candidates to be selected for inclusion in $H_o$ on the next iteration.

---

[2] The "intensity-based" terminology does not relate to "intensification methods" as used in metaheuristic algorithms. However, the "diversity-based" terminology carries an association with "diversification methods" used in such algorithms.

By the sequence of updating $H_o$, the value MinD(i) identified below is equivalent to the value defined as $MinD(i) = Min(d(i,i(1)), \ldots, d(i,i(k_o - 1)))$.

**Simple Diversity-Based Starting Method**
*Initialization*: Choose an arbitrary point $i \in N$, set $i(1) = i$, $k_o = 1$, $H_o = \{i(1)\}$ and $I_o = N \setminus \{i(1)\}$.

**Core Loop**
*For* $k_o = 2$ *to* k:
    Identify $MinD(i) = Min(d(i,j): j \in H_o)$ for each $i \in I_o$.
    Choose $i' = \arg \max(MinD(i): i \in I_o)$ and set $i(k_o) = i'$, $H_o := H_o \cup \{i'\}$ and $I_o := I_o \setminus \{i'\}$ [3]
*EndFor*

The final set $H_o = \{i(k_o): k_o = 1 \text{ to } k\}$ becomes the set of starting points for the K-MinMax or the K-MinSum algorithm. The computational complexity of this method is $O(n \cdot k^2)$, since we examine O(n) points on each of the k – 1 iterations from $k_o = 2$ to k, and examine the distances from each point to $k_o - 1$ other points.

The original proposal for the preceding method (in a different context) gave extended tie-breaking rules for choosing $i(k_o)$ where multiple points qualified to be a point i' satisfying $MinD(i') = Max(MinD(i))$. Instead of employing such rules, we now propose a refinement of the Simple Diversity-Based Starting Method.

The underlying idea is to consider the maximum of the smallest $MinD(i(k_o))$ value for the elements $i(k_o)$ in $H_o = \{i(k_o): k_o = 1 \text{ to } k\}$ to be an indicator of the quality of $H_o$. Due to the construction of $H_o$, this maximum $MinD(i(k_o))$ value results when $i(k_o) = i(k)$, the last element added to $H_o$. Hence, we adopt the goal of maximizing MinD(i(k)), and re-run the algorithm by allowing it to choose this last element i(k) as the new starting element i(1). This approach is based on the supposition that if element i(k) (which is a limiting element in determining the quality of the current $H_o$) belongs to a better set $H_o$, then we increase the possibility of finding such a better $H_o$ by making i(k) the first element of this set, to increase the distance of other points from it. At the conclusion of the execution that has specified the old i(k) to be the new

---
[3] The MinD(i) values used in these computations need not be stored as a vector indexed by i, but can generated and compared sequentially.

i(1), we therefore check whether the resulting final MaxMin distance MinD(i(k)) exceeds this distance on the previous execution (denoted PreviousMaxMin). If so, the maximization objective value is improving and we re-run the algorithm by again setting i(1) = i(k) for this new i(k). The method stops when MinD(i(k)) < PreviousMaxMin, at which point the $H_o$ that was generated at the end of the previous execution (denoted Previous$H_o$) is recovered, thus picking a local optimum for maximizing MinD(i(k)). We refer below to the modified version of the previous Core Loop that handles this objective as the *Core Algorithm*.

**Refined Diversity-Based Starting Method**
*Initialization*: Choose an arbitrary point i ∈ N, set i(1) = i, $k_o$ = 1, $H_o$ = {i(1)} and $I_o$ = N \ {i(1)}. Set PreviousMaxMin = − *Large* and CurrentMaxMin = 0.

**Core Algorithm**
While CurrentMaxMin > PreviousMaxMin
    *For* $k_o$ = 2 *to* k:
        Identify MinD(i) = Min(d(i,j): j ∈ $H_o$) for each i ∈ $I_o$
        Choose i' = arg max(MinD(i): i ∈ $I_o$) and set i($k_o$) = i', $H_o$ := $H_o$ ∪ {i'} and
            $I_o$ := $I_o$ \ {i'}
    *EndFor*
    CurrentMaxMin = MinD(i')
    *If* (CurrentMaxMin ≤ PreviousMaxMin) *then*
        *Break*[4]
    *Else*
        PreviousMaxMin = CurrentMaxMin
        Previous$H_o$ = $H_o$
        i(1) = i'; $H_o$ = {i(1)}; $I_o$ = N \ {i(1)}
    *Endif*
*EndWhile*

The foregoing approach can also choose an element of $H_o$ other than the last to become the new i(1) (for example, selecting a "middle element" i($k_o$) from the sequence, obtained by rounding $k_o$ = (k − 1)/2 to an integer). However, the stopping criterion should continue to be defined in reference to a local maximum for MinD(i(k)), the last element of $H_o$. Hence the only change in the foregoing pseudocode is to select a different element than the last element i(k) in $H_o$ to become the new i(1). The computational complexity of the procedure remains $O(n·k^2)$ provided a constant limit (e.g., 5 or 10) is placed on the number of executions allowed in searching for a locally maximum MinD(i(k)) value.

The value MinD(i(k)) for the last element of $H_o$ can also serve another function, since it gives an estimate of the least distance for separating the PC-centroids. If this estimated distance is small we may conclude there is a significant chance that k has been chosen too large. Moreover, a process of monitoring the successive $MinD(i(k_o))$ values as $k_o$ ranges upward from 2 can disclose $k_o$ values where $MinD(i(k_o))$ abruptly decreases or falls below a minimum desired distance separating the clusters generated.

---
[4] The instruction *Break* means to exit the "current loop" (the innermost loop the instruction is embedded in).

The preceding Refined Diversity-Based method can be further refined by not waiting until $k_o = k$ to apply it, but instead by selecting a value $kCheck < k$ which is treated temporarily as if it were the final $k_o$. In this way, the objective of maximizing $MinD(i(k))$ can be additionally pursued at an earlier stage of building the set $H_o$, thus potentially affording greater leeway for selecting subsequent elements to add to $H_o$ that will keep the MaxMin distance high.

This refinement may be executed for multiple (successively larger) values of kCheck, but we indicate the form of the method for choosing only a single kCheck value. Thus we first iterate from $k_o = 2$ to kCheck, and then after finding a local maximum for the last $MinD(i')$ (for $i' = i(kCheck)$), we continue iterating from $k_o = kCheck + 1$ to k to complete the method (concluding by seeking a maximum value for $MinD(i')$ for $k_o = k$).[5] We call this approach the Compound Diversity-Based Starting Method which we organize to always iterate over $k_o$ from kFirst to kCheck instead of over 2 to k.[6]

**Compound Diversity-Based Starting Method**
*Initialization*: Choose $k' \in [2, k-1]$. Choose an arbitrary point $i \in N$, set $k_o = 1$, $i(1) = i$, $H_o = \{i(1)\}$ and $I_o = N \setminus \{i(1)\}$. Set PreviousMaxMin $= -Large$, CurrentMaxMin $= 0$ and kFirst $= 2$.

While kFirst $\leq$ k
    **Compound Core Algorithm**
    While CurrentMaxMin $>$ PreviousMaxMin
        *For* $k_o =$ kFirst *to* kCheck
            Identify $MinD(i) = Min(d(i,j): j \in H_o)$ for each $i \in I_o$
            Choose $i' = \arg \max(MinD(i): i \in I_o)$ and set $i(k_o) = i'$, $H_o := H_o \cup \{i'\}$ and
                $I_o := I_o \setminus \{i'\}$
        *EndFor*
        CurrentMaxMin $= MinD(i')$
        *If* (CurrentMaxMin $\leq$ PreviousMaxMin) *then*
            Break
        *Else*
            PreviousMaxMin $=$ CurrentMaxMin
            PreviousH$_o$ $= H_o$
            $i(1) = i'$; $H_o = \{i(1)\}$; $I_o = N \setminus \{i(1)\}$
            kFirst $= 2$ (to reset kFirst after having assigned it the value

---

[5] An interesting small value of k' to initiate such an approach is 3.
[6] This permits starting with kFirst $= 2$ and later increasing kFirst to kCheck $+ 1$ accompanied by resetting kCheck $=$ k for the final execution.

        kCheck + 1 following the first execution of the Compound Core Algorithm)
   *Endif*
  *EndWhile*
  (End Compound Core Algorithm)
  *If* kCheck < k *then*
    kFirst = kCheck + 1
    kCheck = k
    PreviousMaxMin = $-$ *Large*; CurrentMaxMin = 0
  *Elseif* kCheck = k then
    *Break*
  *Endif*
*EndWhile*

We consider one last type of diversity-based starting method, called the Targeted Diversity-Based Starting Method, which instead of seeking to maximize MinD(i) at each iteration selects a target value T for MinD(i) and identifies Deviation(i) = |MinD(i) $-$ T| for the goal of minimizing this latter deviation. An appropriate value for T may be determined by first executing one of the preceding diversity-based starting methods and then selecting T to be final the mean or the median of the MinD(i($k_o$)) values generated as $k_o$ ranges from 2 to k (noting that no MinD(i($k_o$)) value exists for $k_o$ = 1), This targeted diversity-based approach will typically yield the same choice of i' = i(k) as by the MaxMin choice of the preceding methods on the final iteration for $k_o$ = k.

We identify a simple instance of this method analogous to the Simple Diversity-Based Starting Method described earlier.

**Simple Targeted Diversity-Based Starting Method**
*Initialization*: Choose an arbitrary point i $\in$ N, set i(1) = i, $k_o$ = 1, $H_o$ = {i(1)} and $I_o$ = N \ {i(1)}.

*For* $k_o$ = 2 *to* k:
  Identify MinD(i) = Min(d(i,j): j $\in$ $H_o$) and Deviation(i) = |MinD(i) $-$ T| for each i $\in$ $I_o$
  Choose i' = arg min(Deviation(i): i $\in$ $I_o$) and set i($k_o$) = i', $H_o$ := $H_o$ $\cup$ {i'} and
    $I_o$ := $I_o$ \ {i'}[7]
*EndFor*

As with the previous diversity-based starting methods, the Targeted Diversity-Based Starting Method can be embodied in refined versions that identify different initial choices of i(1) for restarting the method, continuing in this case as long as the T value computed at the end of each execution is increasing, to yield a local maximum relative to the rule for selecting T. A special variant of the Targeted Diversity-Based Starting Method designed to create a more compact set of diverse points is described in Section A of the Appendix.

---
[7] As in the case of the MinD(i) values, the Deviation(i) values can be identified sequentially and do not need to be stored as vectors indexed by i.

## 7. Primary Intensity-Based Starting Methods

The intensity-based starting methods exploit the structure of the MinMax and MinSum-centroids to generate a starting collection of clusters where the current cluster at each iteration is the best one possible, subject to the elements available for generating a cluster and the chosen value for the cluster size at that iteration. Rather than identifying starting points for the set N(K) in Step 0, the intensity-based methods generate a starting collection of clusters C(h) and the associated set N*(K) of MinMax-centroids or MinSum-centroids to be passed to Step 2, thereby bypassing the first execution of Steps 0 and 1.

These intensity-based procedures require $O(n^2 \cdot (k + \log n))$ effort in comparison with the $O(n \cdot k^2)$ effort of the diversity-based algorithms of Section 6, which should be taken into account in considering their use for large problems. However, the ability of the intensity-based methods to select the best cluster of a given size from the collection of elements available affords an opportunity to reduce the number of iterations consumed by the K-MinMax and K-MinSum algorithms and thereby compensate for the time invested in obtaining the initial clusters.

We employ the following notation and conventions which apply to the adaptive intensity-based methods as well as to the primary methods.

*Notation and Conventions*

$N_o$ = a subset of N consisting of all points at a particular stage of construction available to belong to clusters C(h) not yet created. (Hence $N_o$ begins equal to N, and then shrinks as elements are removed to populate each cluster C(h) as it is created.) We let $n_o = |N_o|$ and to begin take $N_o = N$.

$\delta(i)$ = a logical "indicator variable," where $\delta(i) = $ *True* if $i \in N_o$ and *False* otherwise. (The use of this indicator variable allows the algorithms that follow to be expressed in a convenient form, although more efficient versions of the algorithms may be produced by reference to doubly linked lists.)

$k_o$ = the number of clusters (and hence the number of MinMax-centroids) that remain to be generated, starting at $k_o = k$ and decreasing $k_o$ by 1 at each iteration until reaching 1. Each iteration is identified by its $k_o$ value.

$C(k_o)$ = the cluster generated at iteration $k_o$.

MMCentroid($k_o$) = the MinMax-centroid i* for $C(k_o)$.

ClusterSize = the targeted number of elements $|C(k_o)|$ in the cluster $C(k_o)$. In the simple (first) version of the algorithm described below, ClusterSize receives a value that is (approximately) the same for each $k_o$, while in the succeeding more advanced version ClusterSize varies adaptively for different $k_o$ values.

ScanSize = the number of distances $d(i,j)$ scanned from element i at iteration $k_o$. Since element i itself will belong to the cluster $C(k_o)$ to be generated, ScanSize = ClusterSize − 1.

We determine bounds MinSize and MaxSize on ClusterSize as follows. At least one of the $k_o$ clusters remaining to be created must have a size that equals or exceeds the average number of elements $n_o/k_o$ that can be assigned to these clusters. We adopt the goal of insuring the current cluster $C(k_o)$ will attain this size and hence set

$$\text{MinSize} = \lceil n_o/k_o \rceil$$

(where $\lceil \ \rceil$ is the roof function that rounds fractional values upward). In the version of the algorithm that only allows a single value for the size of $C(k_o)$, we choose ClusterSize = MinSize. We observe that this choice is highly natural for applications such as Cao and Glover (2010) where it is highly desirable to have clusters that are all approximately of the same size.

For the adaptive version of the algorithm we allow ClusterSize to vary between MinSize and an upper bound MaxSize. Let GlobalMinSize be a constant denoting the fewest number of elements permitted in any cluster (set externally, independent of the value MinSize). (For our following purposes, it is useful to choose GlobalMinSize $\geq 2$.) Then on the current iteration, with $k_o$ clusters remaining to be constructed, the maximum number of elements permitted in the current cluster $C(k_o)$ is

$$\text{MaxSize} = n_o - \text{GlobalMinSize} \cdot (k_o - 1)$$

Correspondingly, we set the lower and upper bounds on the number of elements scanned from any given point i by MinScan = MinSize − 1 and MaxScan = MaxSize − 1.

Finally, all of the intensity-based methods are based on a preprocessing step that creates an ordered set of indexes $i(1), i(2), \ldots, i(n-1)$ for each $i \in N$ to sequence the distances $d(i,j)$ for each $i \in N \setminus \{i\}$ in ascending order so that

$$d(i,i(1)) \leq d(i,i(2)) \ldots \leq d(i,i(n-1))$$

This preliminary ordering is potentially the most expensive part of the intensity-based methods, with a time complexity of $O(n^2 \cdot \log n)$. Remaining operations of the intensity-based methods are $O(n^2 \cdot k)$ which makes the total complexity $O(n^2 \cdot (k + \log n))$ which may reduce to the complexity of creating the preliminary ordering if $\log n > k$. An exception occurs if the distances $d(i,j)$ are integers and for each i, $\text{Max}(d(i,j): j \in N) - \text{Min}(d(i,j): j \in N)$ has an $O(n)$ range, in which case a properly constructed bucket sort can sequence the distances for all $i \in N$ in $O(n^2)$ time, and the $O(n^2 \cdot k)$ complexity dominates.

It is to be noted that the preliminary ordering needs to be done only once, and its information can be re-used in refined versions of the algorithms subsequently described and also re-used to generate K-MinMax and K-MinSum clusters for different values of k.

Since the intensity-based methods differ for the K-MinMax and the K-MinSum algorithms, we begin by introducing these starting methods for the K-MinMax setting, starting with the simpler "primary" version that selects the single value ScanSize = MinScan.

**Primary MinMax Starting Method**

*Initialization*: Begin with the ordered indexes i(1), i(2), …, i(n-1) for each $i \in N$ and set $N_o = N$, $n_o = n$ and $\delta(j) = True$, $j \in N$.

For $k_o = k$ to 1,
    ClusterSize = $\lceil n_o / k_o \rceil$ (ClusterSize = MinSize)
    ScanSize = ClusterSize – 1
    Execute the **MinMax Distance Algorithm** (described below)
        (At the conclusion: i* = arg min(Max (d(i,j): $j \in C$) (= MMSpan(C)) over all sets
        $C \subset N_o$ such that |C| = ClusterSize.)

    **Generate $C(k_o)$ and update the set $N_o$**
    (This step re-executes the relevant part of the MinMax Distance Algorithm for i = i* to
        determine the set $C = C(k_o)$ associated with i*.)
    $C(k_o) = \emptyset$
    Scan = 0
    *For* s = 1 *to* n – 1
        j = i*(s)
        *If* ($\delta(j)$ = True) *then* (j is an element of $N_o \setminus \{i\}$
            Scan := Scan + 1
            $N_o := N_o \setminus \{j\}$
            $C(k_o) = C(k_o) \cup \{j\}$
            $\delta(j) = False$
            *If* (Scan = ScanSize) *Break*
        *Endif*
    *Endfor*
    $N_o := N_o \setminus \{i^*\}$
    $C(k_o) = C(k_o) \cup \{i^*\}$
    $\delta(i^*) = False$
    MMCentroid($k_o$) = i*
    $n_o := n_o - $ ClusterSize

*Endfor*

On the final iteration of the Primary Algorithm when $k_o = 1$, the step of generating $C(k_o)$ can be shortcut by simply setting $C(1) = N_o$ and $MMCentroid(1) = i^*$ (and there is no need to update $N_o$ because it will become empty). This ability to shortcut the final update of $C(k_o)$ for $k_o = 1$ also holds for all intensity-based starting methods subsequently discussed.

The MinMax-centroids of $MMCentroid(k_o)$, $k_o = 1, \ldots, k$, generated by the foregoing method give the initial set of points $N^*(K) = (i(h), h \in K\}$ that can be passed directly to Step 2 of the K-MinMax Algorithm along with the associated clusters $C(k_o)$. This likewise is true of all intensity-based methods subsequently discussed (referring to the array $MSCentroid(k_o)$ in the case of the K-MinSum algorithm).

The internal MinMax Distance algorithm executed within the Primary Algorithm, described next, is understood to share its arrays and values with the Primary Algorithm.

**MinMax Distance Algorithm**
*Conventions:*
$i^*$ identifies a point i that yields a MinMax-centroid for a "best" C of size ClusterSize in $N_o$.
BestDistance = the value $Min((Max (d(i^*,j): j \in C): C \subset N_o)$ that gives $MMSpan(C)$ for this
     "best" C.
*Large* = a large positive number

*Initialization:*
BestDistance = *Large*

For $i \in N_o$
    Scan = 0 (the number of elements j scanned from i)
    *For* s = 1 *to* n – 1
        j = i(s)
        *If* ($\delta(j)$ = True) *then* (j is an element of $N_o \setminus \{i\}$)
            Scan := Scan + 1
            *If* (Scan = ScanSize) *Break*
        *Endif*
    *Endfor*
    *If* (d(i,j) < BestDistance) *then*
        $i^* = i$
        BestDistance = d(i,j)
    *Endif*
*Endfor*

The rationale underlying the foregoing method may be expressed as follows. Let v = ClusterSize and $n_o = |N_o|$, for ClusterSize and $N_o$ determined at iteration $k_o$, and let CSet = $\{C \subset N_o: |C| = v\}$ denote the set of all clusters C in $N_o$ of size v. Assume at each iteration $k_o$ we seek to identify a "best" cluster $C_{best}$ that minimizes MMSpan(C) over the clusters C in CSet. Evidently, one way to do this, which we would hope to improve upon, is to examine the clusters in CSet (where $|CSet| = n_o!/v!(n_o - v)!$) and then determine the MinMax-centroid i' of each C in CSet to compute MMSpan(C) = Max(d(i',j): j ∈ C \ {i'}). The following result establishes the ability of the Primary Starting Method to achieve the desired outcome with vastly less effort.

**Theorem 1.** The cluster $C(k_o)$ and the associated point i* identified at the conclusion of iteration $k_o$ of the Primary Starting Method respectively qualify to be $C_{best}$ and its MinMax-centroid.

*Proof*: Let $i_o(t)$, $t = 1, \ldots, n_o - 1$ identify the subsequence of i(s), $s = 1, \ldots, n - 1$ such that $\delta(j) = \textit{True}$ for j = i(s). Hence, defining $j_t = i_o(t)$ (with the identity of i implicit), the ordering of distances d(i,j) for point i can be written $d(i,j_1) \leq d(i,j_2) \leq \ldots \leq d(i,j_{no-1})$. Consider the cluster $C_i$ consisting of i and the points $j_t$ for t = 1 to v – 1. (Note that t corresponds to successive values of Scan, and v – 1 = ScanSize.) It is possible that i is not a MinMax-centroid of $C_i$ and hence the distance $d(i,j_{v-1})$ may be larger than MMSpan($C_i$). Nevertheless, for a point i* that qualifies as a MinMax-centroid of $C_{best}$ (at least one such i* must exist) there can be no better candidate for $C_{best}$ than the set $C_{i*}$, and the distance $d(i*,j_{v-1})$ must equal MMSpan($C_{best}$). Therefore, letting BestDistance denote the minimum of the $d(i,j_{v-1})$ values over all points i examined so far, the Primary MinMax Starting Method goes through the successive indexes $j_t$, t = 1 to v – 1, for each i ∈ $N_o$ (where $j_t$ depends on i) and stops at t = v – 1 to check whether $d(i,j_{v-1})$ < BestDistance. If the inequality holds, then BestDistance is updated to be the current $d(i,j_{v-1})$ and i* is recorded as i, which verifies the validity of the Starting Method and establishes the claim of the theorem. □

Note the foregoing proof also shows it would be possible to interrupt the sequence of t values for a given i ∈ $N_o$ if $d(i,j_t) \geq$ BestDistance, and thus break from the "For loop" at this point. The foregoing analysis also makes it clear that the Primary MinMax Starting Method succeeds in finding a best C and its MinMax-centroid for a given $N_o$ in $O(n_o \cdot n)$ time. (This time could be reduced to $O(n_o \cdot v)$ if the sequence $i_o(t)$ were used directly instead of relying on the check $\delta(j) = \textit{True}$ for j = i(s)). For the iterations for $k_o = k$ to 1 the total effort is therefore bounded above by $O(n^2 \cdot k)$, as remarked earlier.

We are motivated to improve upon the outcome obtained by the Primary Starting Method by accounting for the fact that an ideal collection of clusters may well have somewhat different numbers of elements in different clusters. One way to pursue such an improvement is to look beyond the end of the execution of the Primary Starting Method to identify values for ClusterSize = ClusterSize($k_o$), for $k_o$ = 1 to k which are then used to launch a second execution of the Starting Method. We consider two such approaches next.

*Refinement to Choose Different ClusterSize values.*

*Approach 1*: After applying the Primary Starting Method (to bypass Steps 0 and 1 of the K-MinMax algorithm), perform the first execution of Step 2 to determine a new collection of clusters C(h), h ∈ K. Index these clusters so the values |C(h)| are in descending order for h = k to

1 and set ClusterSize(h) = |C(h)|. Finally, rerun the Primary Starting Method by setting ScanSize = ClusterSize($k_o$) − 1 in the "For loop" from $k_o$ = k to 1. We choose to put the ClusterSize($k_o$) values in descending order as $k_o$ goes from k to 1 by noting that $N_o$ will be larger, and hence will offer greater latitude for choosing C($k_o$), for larger values of k. (This approach may be shortened by interrupting the Primary Starting Method after executing it for any value $k_o$* and reassigning the elements of N \ $N_o$ in Step 2 to obtain a subset of clusters C(h) for h = k to $k_o$*. The indexes of clusters are then arranged as indicated above to produce descending ClusterSize($k_o$) for $k_o$ = k to $k_o$*. Remaining ClusterSize values for $k_o$ = $k_o$*− 1 to 1 can be determined by ClusterSize = ⌈$n_o$/$k_o$⌉ exactly as before.)

*Approach 2*: This approach is the same as the first, except that it waits until a later iteration of the K-MinMax Algorithm to select and order the collection of clusters C(h) at Step 2. If the approach waits until the final iteration, the overall effort of generating the final clusters may be expected to be less than twice the amount of the first execution of the K-MinMax Algorithm, because effort is saved on the second execution by not having to re-do the preprocessing step that determines the sequence (i(s): s = 1, …, n-1) for each i ∈ N. In addition, fewer iterations may be required on the second execution due to the chance of generating a better set of starting clusters.

The Adaptive Intensity-Based Starting Methods described in Section 8 give a different way to account for the fact that the best clusters C(h) can vary in size.

**Primary Intensity-Based Method for the K-MinSum Problem**

We organize the Primary Intensity-Based Starting Method for the K-MinSum Algorithm in the same manner as for the K-MinMax Algorithm. This method is quite similar to the Primary MinMax Starting Method, the chief difference being to replace MaxDist(i) with SumDist(i), and to replace the list MMCentroid($k_o$) with a corresponding list MSCentroid($k_o$) to identify the MinSum-centroids generated. To avoid ambiguity concerning this correspondence, and to clarify how the MinSum process differs from the MinMax process, we identify the Primary MinSum Starting Method in Section B of the Appendix.

## 8. Adaptive MinMax and MinSum Starting Methods

As already noted, the adaptive starting methods for the K-MinMax and K-MinSum algorithms add a layer of sophistication beyond that of the primary starting methods, in order to respond to the challenge of determining how to select varying cluster sizes in the absence of prior knowledge about the ideal sizes. We begin by considering an adaptive method for the K-MinMax algorithm.

## 8.1 Foundation of the Adaptive MinMax Starting Method

The Adaptive MinMax method has two phases. Phase 1 resembles the Primary MinMax method by identifying a BestDistance value over all $i \in N_o$ for ClusterSize = MinSize. In addition, it identifies a value DistanceLimit that limits the largest distance $d(i,j)$ based on the successive differences in the ascending $d(i,j)$ values that are used to compute BestDistance.

Phase 2 then makes use of the information generated in Phase 1 by examining each $i \in N_o$ and all elements j in the set $J(i) = \{j \in N_o \setminus \{i\}: d(i,j) \leq DistanceLimit)$ to identify a set of "best points" i as candidates for the MinMax-centroid i*. This set of points is defined by BestSet = $\{i \in N_o: |J(i)| = MaxJ\}$ where $MaxJ = Max(|J(i)|: i \in N_o)$. Finally, to choose among elements in BestSet, the method conputes the value $D(i) = Max(d(i,j): d(i,j) \leq DistanceLimit)$ and selects the element i* that gives.

$$i^* = \arg \min \ (D(i): i \in BestSet)$$

This construction insures i* will be a MinMax-centroid over the set $C = J(i^*) \cup \{i^*\}$ and, as in the case of the Primary MinMax method, will identify a best such set among those of the same size. Since the adaptively determined cluster C can contain somewhat more than MinSize elements, the upper limit MaxSize is imposed by the calculation indicated earlier as a safeguard against the possibility that $|C|$ may become too large.

If the procedure is modified to permit MaxSize to receive a value as large as $|N_o|$, then the current cluster could potentially absorb all of $N_o$, and under such a condition the method will terminate with fewer than k clusters.[8] Such a variation can be of interest for enabling the adaptive starting method to propose a new (smaller) value for k. An evaluation criterion to verify when fewer clusters can be appropriate is suggested in Section 9.

In spite of the more advanced character of the adaptive starting method, the use of the ordering (i(s): s = 1, …, n-1) makes it possible to identify the sets J(i) and values D(i) in Phase 2 of the adaptive method implicitly, rather than explicitly, as part of the process of identifying i*.

**Adaptive MinMax Starting Method**
*Initialization*: Begin with the ordered indexes i(1), i(2), …, i(n-1) for each $i \in N$ and set $N_o = N$, $n_o = n$ and $\delta(j) = \text{True}$, $j \in N$.

For $k_o = k$ to 1
    MinSize = $\lceil n_o / k_o \rceil$
    MaxSize = $n_o - \text{MinSize} \cdot (k_o - 1)$
    MinScan = MinSize − 1

---

[8] An exception occurs when $k_o = 1$, at which point all of $N_o$ is absorbed in any case.

    MaxScan = MaxSize − 1

    Execute the **Phase 1 MinMax Algorithm** (to identify BestDistance)
    Execute the **Phase 2 MinMax Algorithm** (to identify i*)
    **Generate C($k_o$) and update the set $N_o$**
    (This portion of the Adaptive MinMax Method is identical to that of the Primary
    MinMax Method, except that "*If* (Scan = BestScan) *Break*" replaces
    "*If* (Scan = ScanSize) *Break*".)
*Endfor*

Next we describe the Phase 1 and Phase 2 algorithms embedded in the Adaptive MinMax method. We employ a bare minimum of descriptive comments, since the mnemonic names of the variables and the instructions of the algorithm should make the interpretation clear.

**Phase 1 MinMax Algorithm**
*Assumption:* $N_o$ contains at least 2 elements
BestDistance = *Large*
BestMinGap = 0

For i ∈ $N_o$
    MinGap = *Large*
    Sum = 0
    SumGap = 0
    Scan = 0
    *For* s = 1 *to* n − 1
        j = i(s)
        *If* (δ(j) = *True*) *then*
            Scan := Scan + 1
            Distance = d(i,j)
            Sum := Sum + Distance
            *If* (Scan > 1) *then*
                Gap = Distance − PreviousDistance
                SumGap = SumGap + Gap
                MinGap = Min(Gap,MinGap)
            *Endif*
            *If* (Scan = MinScan) *then*
                s(i) = s (initialize Phase 2 at s = s(i))
                *Break*
            *Endif*
            PreviousDistance = Distance
        *Endif*
    *Endfor*
    *If* ((Distance < BestDistance)

```
                or ((Distance = BestDistance) and (MinGap > BestMinGap)) ) then
                    BestDistance = Distance
                    BestMinGap = MinGap
                    BestSumGap = SumGap
            Endif
    Endfor
    BestMeanGap = BestSumGap/(MinScan − 1)
    TargetGap = λ·BestMeanGap + (1 − λ)·BestMinGap)  (suggested default, λ = .3)
    (The following value is inherited in Phase 2 as a limit for i ∈ N_o)
    DistanceLimit = BestDistance + TargetGap
```

**Phase 2 Adaptive MinMax Algorithm**

*Conventions:*
DistanceLimit = the largest distance allowed for d(i,j) when scanning from element i.
BestScan = the value of Scan that produces the best candidate for i*.
*Inherited from Phase 1:* s(i), i ∈ $N_o$, and DistanceLimit
*Initialization:*
BestDistance = *Large*
BestScan = MinScan − 1

```
For i ∈ N_o
    Scan = MinScan − 1 (Scan is immediately incremented to MinScan in the next loop)
    For s = s(i) to n − 1
        j = i(s)
        If (δ(j) = True) then  (automatically true for s = s(i))
            Scan := Scan + 1 (reaches MinScan on first execution, when s = s(i))
            If (d(i,j) > DistanceLimit) then
                Scan := Scan − 1
                Break
            Endif)
            Distance = d(i,j)
            If (Scan = MaxScan) Break
        Endif
    Endfor
    If (Scan ≥ MinScan) then  (i is a candidate for i*)
        If ((Scan > BestScan)
            or ((Scan = BestScan) and (Distance < BestDistance))) then
            i* = i
            BestScan = Scan
            BestDistance = Distance
        Endif
    Endif
Endfor
```

We observe that Approaches 1 and 2 of the "Refinement to Choose Different ClusterSize Values" for the Primary MinMax Starting Method can also be used with the Adaptive MinMax Starting Method, where upon restarting the K-MinMax algorithm for the second pass the primary starting method is used together with setting ClusterSize = ClusterSize($k_o$).

The justification of the Adaptive MinMax Starting Method is provided in Theorem 2, following. As in the discussion that precedes Theorem 1, let v = ClusterSize , where in this case ClusterSize is not specified in advance but is determined at iteration $k_o$ to be the value BestScan + 1 for BestScan at the conclusion of this iteration. As before, let CSet = {C ⊂ $N_o$: |C| = v}. Now at each iteration $k_o$ we seek to identify a "best" cluster $C_{best}$ that minimizes MMSpan(C) over the clusters C in CSet, where v is not fixed at the start of iteration $k_o$. Subject to the restriction Scan ≤ MaxScan, which implies v ≤ MaxScan + 1, we undertake to make v as large as possible. Because of this dependency on the variable value of v, we refer to $C_{best}$ as $C_{best}(v)$ and to the best value of v as v*, hence identifying the best cluster C finally obtained as $C_{best}(v^*)$.

**Theorem 2.** The cluster C($k_o$) and the associated point i* identified at the conclusion of iteration $k_o$ of the Adaptive MinMax Starting Method respectively qualify to be $C_{best}(v^*)$ and its MinMax-centroid.

*Proof*: As in the proof of Theorem 1, let $i_o(t)$, t = 1, …, $n_o$ – 1 identify the subsequence of i(s), s = 1, …, n – 1 such that δ(j) = *True* for j = i(s). Hence, defining $j_t = i_o(t)$ (with the identity of i implicit), the ordering of distances d(i,j) for point i can be written d(i,$j_1$) ≤ d(i,$j_2$) ≤ … ≤ d(i,$j_{no-1}$). Now the cluster $C_i$ consisting of i and the points $j_t$ for t = 1 to v – 1 depends on v as a variable rather than as a constant. The first priority is to make v as large as possible, subject to requiring v ≤ MaxScan + 1, and more particularly subject to requiring d(i,$j_{v-1}$) ≤ DistanceLimit where DistanceLimit is determined by BestDistance and TargetGap computed in Phase 1 according to the choice of the parameter λ. The best value v* for v is the value BestScan + 1 for BestScan as updated in Phase 2, whenever BestScan can be increased subject to BestScan ≤ MaxScan. The final "*If … then*" instruction in Phase 2 first increases BestScan (hence implicitly increases v*) when possible, and otherwise decreases the value BestDistance when BestScan remains unchanged. By this layer of priorities, the method always first increases v*, and for v* unchanged selects the minimum value d(i,$j_{v^*-1}$) for BestDistance. Consequently, by the same reasoning as in the proof of Theorem 1, it follows that ultimately $C_{best}(v^*)$ identifies a cluster that is the best for the largest admissible value v* for v, and thus qualifies to be treated as the cluster denoted $C_{best}$ in Theorem 1 with i* constituting its associated MinMax-centroid. □

The analysis of the complexity of the Primary MinMax Starting Method applies as well to the Adaptive MinMax Starting Method in spite of its more elaborate form.

This result carries over with the replacement of "MinMax" by "MinSum" and "MMCentroid" by "MSCentroid" to become applicable to the corresponding starting method for the K-MinSum problem.

## 8.2 The Adaptive MinSum Starting Method

The Adaptive MinSum Starting Method follows a pattern similar to that of the Adaptive MinMax Starting Method with some minor departures. As in the case of the Primary MinSum Starting Method, we include the pseudo-code for the Adaptive MinSum method in Section B of the Appendix to clearly disclose the details where the two methods differ.

## 9. A Regret-Threshold PC Algorithm and a Diversified Restarting Method

We conclude by proposing a modification of the K-PC Algorithm called a *Regret-Threshold PC Algorithm* that applies both to the K-MinMax and K-MinSum algorithms, and which can also be used to modify the K-Means Algorithm by replacing references to the pseudo-centroid with references to the ordinary centroid.

The strategy of the Regret-Threshold algorithm is to form a candidate list consisting of the elements that are reassigned to new clusters in Step 2 of the K-PC Algorithm and to permit only elements belonging to a subset of this candidate list to be reassigned. The elements eligible for reassignment are determined by a threshold value T which is based on a regret measure, as follows.

Referring to Step 2 of the K-PC Algorithm, let CandidateList = {j ∈ N: j qualifies to be reassigned to a new cluster C(h)}, and for j ∈ CandidateList, let:
    i' = Assign(j) denote the pseudo-centroid of the cluster C(h') that j is currently assigned to
    i" = ReAssign(j) denote the pseudo-centroid of new cluster C(h") to which j would be
        reassigned by Step 2.
    ADist(j) = d(i',j) denote the *assign distance* for j
    RDist(j) = d(i",j) denote the *reassign distance* for j.
    Regret(j) = ADist(j) − RDist(j)

Hence, the larger the value of Regret(j), the greater is the regret for not reassigning j to the new pseudo-centroid ReAssign(j) (and hence to the associated cluster C(h")) in Step 2. These definitions equally apply in the situation where multiple points may be recorded as pseudo-centroids of a cluster C(h), as represented by the set C*(h).

We then use the threshold T by identifying a selected number r of the "largest regret" elements in CandidateList, as where r is determined to admit some fraction F of these elements for consideration. As a simple example, F may be chosen from the interval [.05, .2] subject to assuring that at least 1 element passes the threshold. When F = 1, this approach is the same as the K-PC approach. (Instead of making F constant, a natural variant is to begin by selecting F larger and then allowing it to become smaller as the number of iterations grows.)

Upon assigning T a value that will accept these specified elements, the eligible elements to be reassigned in Step 2 are those belonging to

$$\text{SelectList} = \{j \in \text{CandidateList}: \text{Regret}(j) \geq T\}$$

An upper bound may be additionally imposed on the size of SelectList when ties in the regret values would cause SelectList to contain more elements than considered desirable.

The foregoing Regret-Threshold approach may be accompanied by a restarting algorithm to create a new set of initial clusters that differ strategically from the set of clusters produced at the termination of a current execution, as a basis for obtaining a new set of final clusters on an ensuing execution. This algorithm, which we call the Diversified Restart Method, may be used with the earlier K-PC algorithms as well as with the Regret-Threshold approach. Accordingly, we use the term "Clustering Algorithm" to refer to any of these algorithms.

**Diversified Restart Method**

*Terminology*: Let $C(h)$, $h \in K$, denote the final set of clusters produced by the Clustering Algorithm, and let $C^*(h)$ denote the set of pseudo-centroids associated with $C(h)$ and $C^*$ denote the union of the sets $C^*(h)$, $h \in K$. (We allow the option of saving only one element in $C^*(h)$.) For each $j \in N$, let $h = h(j)$ identify the cluster $C(h)$ containing $j$ and denote the pseudo-centroid to which $j$ is assigned by $i = PC_1(j) \in C^*(h)$. (Thus $PC_1(j)$ is the pseudo-centroid in $C^*$ closest to $j$.) Finally, let $d_1(j) = d(j, PC_1(j))$ and let $d_2(j) = \text{Min}(d(j,i): i \in C^* \setminus C^*(h)$ for $h = h(j))\}$. (Hence $d_1(j)$ is the distance from the pseudo-centroid "first closest" to $j$, while $d_2(j)$ is the distance from the pseudo-centroid "second closest" to $j$, restricting attention in the latter case to pseudo-centroids outside of $C^*(h)$.)

1. Beginning from the final set of Clusters $C(h)$, create a new set of clusters (to start another execution of the clustering algorithm) by assigning each $j \in N \setminus C^*$ to the pseudo-centroid $PC_2(j) \in C^*$ to $j$.

2. Restart the Clustering Algorithm with the new set of clusters and executing Step 1 of the Clustering Algorithm to identify new pseudo-centroids for each $C(h)$.

We observe that the first step above associates each $j \in N \setminus C^*$ with a cluster $C(h')$ (containing $PC_2(j)$) that is different from the cluster $C(h)$. An alternative is to reassign only a portion of the points $j \in N \setminus C^*$ to an alternative pseudo-centroid before restarting.

More pronounced forms of diversification can be achieved by redefining $PC_2(j)$ to be the pseudo-centroid that is "third closest" to $j$, or in the extreme to be the pseudo-centroid farthest from $j$. However, according to the types of diversification often favored in metaheuristic methods, the definition of $PC_2(j)$ above is likely to be preferred.

**Alternative Measure for Evaluating the Collection of Clusters**

The definitions underlying the Diversification Method give a natural way to create a function for evaluating the clusters, which also can be used to identify good values for k.

Let $d_o(j) = d_2(j) - d_1(j)$ ($\geq 0$). The magnitude of $d_o(j)$ signals the relative appropriateness of assigning j to a cluster with pseudo-centroid $PC_1(j)$, as opposed to assigning j to some other clusters in the current collection. We can exclude elements of $C^*$ from having an impact by the convention $d_o(j) = 0$ for $j \in C^*$, or we can include them by defining $d_1(j) = 0$ for $j \in C^*$ in the definition of $d_o(j)$.

Then an overall measure of the "quality" of C(h) is given by

$$D_o(h) = \sum(d_o(j): j \in C(h)) \text{ or by } \text{Mean}_o(h) = D_o(h)/|C(h)|.$$

The denominator $|C(h)|$ may be replaced by $|C(h)\backslash C^*(h)|$ if pseudo-centroids are excluded from consideration in defining $d_o(j)$. Finally, by drawing on $D_o(h)$ we may evaluate the entire collection of clusters C(h), $h \in K$ by

$$\text{Value} = \sum (\text{Mean}_o(h): h \in K)$$

Here larger values indicate higher quality.[9]

The Value term may be considered an alternative to the Davies-Bouldin validity index (Davies and Bouldin, 1979) which is frequently used to compare the quality of different collections of clusters. Since *Value* can meaningfully compare cluster collections for different values of k, this term can be used with other approaches such as those suggested in earlier sections for evaluating collections to find a preferred k value.

The $d_o(j)$ quantities used to define $D_o(h)$ and $\text{Mean}_o(h)$ can also serve to create a stochastic variant of the Diversified Restart approach. In this case, probabilities can be computed for reassigning points that are inversely related to the magnitude of the $d_o(j)$ values. Then, as an extension of the variant that only permits a subset of the elements $j \in N \setminus C^*$ to be reassigned, a cutoff value can be established to prevent reassignment of points with $d_o(j)$ values exceeding a specified magnitude, and the probabilities for reassigning remaining elements can be used to accept or reject such reassignments. Such a strategy reflects the heuristic notion that points with the largest $d_o(j)$ values in all likelihood should be maintained in a common cluster.

## 10. Conclusions

The proposals of the preceding sections provide a wide range of strategies for clustering with pseudo-centroid methods, and lay a foundation for studies to compare these methods across alternative classes of clustering problems. Other types of pseudo-centroids, such as a MinProduct-centroid, a MinMedian-centroid, which are defined in the obvious way, give rise to associated K-MinProduct and K-MinMedian clustering algorithms whose forms can be identified by analogy with the K-MinMax and K-MinSum algorithms.[10] Similarly, we can identify K-PC algorithms associated with a MinSpan-centroid for Span = Max − Min, and with various "ratio-based" pseudo-centroids, such as a MinSpanRatio-centroid based on a ratio such as Max/Min (excluding Min = 0, conceiving that two points separated by a 0 distance represent the same point) and a MinSumRatio-centroid based on a ratio such as Sum/Max ratio (or Sum/Median). These variations employing more subtle types of pseudo-centroids also invite study.

---

[9] A variation to accentuate the influence of points with larger $D_o(h)$ values is to define $\text{Mean}_o(h) = D_o(h)^2/|C(h)|$.

Another potential area for exploitation arises by noting that the pseudo-centroid approaches may be contrasted with fuzzy clustering approaches as in Sudha et al. (2012) and imprecise knowledge clustering approaches as in Anwar et al. (1992). An interesting topic for research would be to integrate elements of these latter approaches with the pseudo-centroid approaches.

In recent years a variety of metaheuristic algorithms for clustering have emerged to obtain better outcomes than those produced by classical approaches, which in effect are heuristics for obtaining local optima for associated evaluation functions (though these functions are not always clearly defined). While the results and procedures of this paper are offered independent of the metaheuristic context, we observe that metaheuristic algorithms can be employed to enhance the effectiveness of these clustering approaches in the same way that metaheuristic algorithms are used to enhance the K-Means and K-Medoid approaches. Conversely, there is a great potential to enhance the operation of metaheuristic algorithms by making use of clustering. Proposals of this type have been made, for example, to improve intensification and diversification strategies for tabu search and to refine the rules for selecting solutions to be combined by scatter search and path relinking (which could also apply to solutions combined by genetic algorithms).[11] The notions underlying pseudo-centroid clustering may find applications in these areas as well.

*Compliance with Ethical Standards:*

Funding: This study was not funded.

Conflict of Interest: The author declares that he has no conflict of interest.

Ethical approval: This article does not contain any studies with human or animal participants.

---

[10] The K-MinProduct problem can also be approached via the K-MinSum algorithm by using logarithms. The K-MinMedian problem is not to be confused with the K-median problem, which is based on a different concept relying on coordinate vectors. We can also define a K-MedianMedian algorithm, among other novel variants.

[11] See, for example, Glover (1994, 1997)

# Appendix: Additional Supporting Methods and Pseudocode

## A. Diversity-Based Starting Methods

### 1. A Targeted Diversity-Based Starting Method to Produce More Compact Sets of Points

A more elaborate variant of the Targeted Diversity-Based Starting Method stems from the observation that for some number of iterations, there are likely to be numerous ties for the point $i' = \arg \min(\text{Deviation}(i): i \in I_o)$, that is, there may be many points $i'$ whose minimum distance from the points $j \in H_o$ is close to T (yielding Deviation($i'$) close to 0), and we would prefer to choose among them in a way that keeps the growing collection of points in $H_o$ compact. This potential for ties is particularly evident if we choose T to be the value MinD($i(k)$) which is the final MaxMin value when the last element $i(k_o) = i(k)$ is added to $H_o$ by the Simple Diversity-Based Starting Method. Since the point $i(k)$ lies at a distance at least $T = \text{MinD}(i(k))$ from all other points in $H_o$, selecting $i(k)$ as the starting point $i(1)$ yields numerous points $i$ for which Deviation($i$) = |MinD($i$) − T| is close to 0. As a result, a sequence of points chosen to populate $H_o$ could initially resemble a straight line (and subsequently a bent line with gradual fill-in), whose endpoints get progressively farther away from the points added earlier. Such a "line configuration" would be encouraged, for example, by breaking ties in the choice of $i'$ to favor points farthest from the most recent point $i(k_o)$ added to $H_o$.

We introduce two refinements as a basis for generating a more compact collection of points in $H_o$. First, we identify a target T and a starting point $i(1)$ as follows. Relative to the set $H_o = \{i(k_o): k_o = 1 \text{ to } k\}$ generated from an execution of the Simple Diversity-Based method or one of its refined variants, identify the min distance for each $i = i(k_o) \in H_o$ from the other points in $H_o$, given by $\text{MinD}_o(i) = \text{Min}(d(i,j): j \in H_o \setminus \{i\})$. (This is effectively equivalent to the previous definition of MinD($i$), since the latter is restricted to elements $i$ not in $H_o$.) Then we identify the mean of these MinD$_o$($i$) values given by

$$\text{MeanMinD} = \sum \text{MinD}_o(i): i \in H_o)/k$$

Finally, we select the target T = MeanMinD and pick a point $i^\#$ of $H_o$ to become the new $i(1)$ (to generate a new $H_o$) whose MinD$_o$($i$) value is closest to MeanMinD:

$$i^\# = \arg \min(|\text{MinD}_o(i) - \text{MeanMinD}|: i \in H_o)$$

The choice of T = MeanMinD and $i(1) = i^\#$ to launch the targeted Diversity-Based Starting Method is then accompanied by introducing a tie-breaking rule for choosing $i'$ that minimizes the maximum distance from the points currently in $H_o$.

For greater latitude of choice, we choose a small value $T_o$ and require all qualifying points to belong to the set $I_o' = \{i \in I_o : \text{Deviation}(i) \leq \text{MinDev} + T_o\}$, where we define $\text{MinDev} = \text{Min}(\text{Deviation}(i): i \in I_o)$ (hence all points that satisfy $\text{Deviation}(i) = \text{MinDev}$ qualify to be chosen as i' in the Targeted Diversity-Based method). For example, $T_o$ can be obtained by setting $T_o = f \cdot T$ for is a small non-negative fraction f. (When $f = 0$ and hence $T_o = 0$, the set $I_o'$ consists strictly of the points that qualify to be i' in the Simple Targeted Diversity-Based method.)

Finally we define $\text{MaxD}(i) = \text{Max}(d(i,j): j \in H_o)$ and choose i', the element to be added to $H_o$ at each iteration, by $i' = \arg\min(\text{MaxD}(i): i \in I_o')$, thereby minimizing the maximum distance from points in $H_o$, given that i' lies in $I_o'$. This gives the following method.

**Targeted Diversity-Based Starting Method with Flexible Tie-Breaking**
*Initialization*: Choose the point $i(1) = i^\#$ and the target $T = \text{MeanMinD}$, and set $k_o = 1$, $H_o = \{i(1)\}$ and $I_o = N \setminus \{i(1)\}$.

*For* $k_o = 2$ *to* k:
    Identify $\text{MinD}(i) = \text{Min}(d(i,j): j \in H_o)$, $\text{Deviation}(i) = |\text{MinD}(i) - T|$ for each $i \in I_o$,
        $\text{MinDev} = \text{Min}(\text{Deviation}(i): i \in I_o)$, $I_o' = \{i \in I_o : \text{Deviation}(i) \leq \text{MinDev} + T_o\}$
        and for $i \in I_o'$, identify $\text{MaxD}(i) = \text{Max}(d(i,j): j \in H_o)$.
    Choose $i' = \arg\min(\text{MaxD}(i): i \in I_o')$ and set $i(k_o) = i'$, $H_o := H_o \cup \{i'\}$ and
        $I_o := I_o \setminus \{i'\}$
*EndFor*

This approach likewise can be iterated, by choosing a new the target $T = \text{MeanMinD}$ and starting point $i(1) = i^\#$ from the $H_o$ most recently generated. A natural variant of the method results by replacing $\text{MaxD}(i)$ with $\text{SumD}(i) = \sum(d(i,j): j \in H_o)$.

The earlier Diversity-Based Starting Methods of Section 6 can similarly be modified to generate more compact sets of points by defining $\text{MaxMin} = \text{Max}(\text{MinD}(i): i \in I_o)$ and $I_o' = \{i \in I_o : \text{MinD}(i) \geq \text{MaxMin} - T_o\}$. Then as in the Targeted Tie-Breaking Approach above, for $i \in I_o'$ we identify $\text{MaxD}(i) = \text{Max}(d(i,j): j \in H_o)$ and choose $i' = \arg\min(\text{MaxD}(i): i \in I_o')$ (or alternatively replace $\text{MaxD}(i)$ with $\text{SumD}(i)$). The starting point $i(1) = i^\#$ can also be used, without reference to the target T.

## 2. Successive Elimination Diversity-Based Starting Methods

Another type of diversity-based starting methods employs a successive elimination strategy that shares an organization related to that of the Intensity-Based methods, except that it strictly aims to create a diverse set of points, again denoted by $H_o$, instead of to generate clusters at the same time. The basic strategy is as follows.

*Initialization*: Set $N_o = N$, $n_o = n$ and $H_o = \varnothing$.

*For* $k_o = 1$ *to* k:

    Apply an Elimination Choice Rule (as identified below) to select a point $i' \in N_o$ and let $H_o := H_o \cup \{i'\}$ and $N_o := N_o \backslash \text{Proximity}(i')$, where Proximity($i'$) denotes the set of $\lceil n_o / k_o \rceil$ points $j \in N_o \backslash \{i'\}$ that are closest to i.

*EndFor*

A trivial possibility for the Elimination Choice Rule would be to simply choose $i'$ randomly from $N_o$. More interesting possibilities are as follows, where for each point $i \in N_o$, let Proximity(i) be defined the same as the set Proximity($i'$), i.e., the set of $\lceil n_o / k_o \rceil$ points $j \in N_o \backslash \{i\}$ that are closest to i.

*Elimination Choice Rule*. Let PC(i) = a PC-centroid of Proximity(i) according to the case in question, and let V(i) = Span(Proximity(i)). (For example, in the MinMax case, PC(i) is a point $i^*$ in Proximity(i) whose maximum distance from all other points in Proximity(i) is smallest, and V(i) = MaxDist($i^*$: Proximity(i)).) Then choose $i'$ in $N_o$ to be a point that gives a value of V(i) that is (a) largest, (b) smallest or (c) 'closest to the mean' of such values for $i \in N_o$.

Using an instance (a), (b) or (c) of the preceding rule, an iterated version of the method can be employed by successively restarting again from $k_o = 1$, pre-selecting the value $i'$ for $k_o = 1$ to be the "final $i'$ value" for $k_o = k$ (chosen at the end of the preceding iteration). In this iterated approach, a variation is to draw on ideas related to those of other Diversity-Based methods by re-defining Proximity(i) after the first iteration (over $k_o = 1$ to k) to consist not just of the $\lceil n_o / k_o \rceil$ points $j \in N_o \backslash \{i\}$ that are closest to i, but to encompass all points $j \in N_o$ such that $d(i,j) \leq U$, where U is the mean or maximum $d(i',j)$ value for $j \in \text{Proximity}(i')$ over $i' \in H_o$ on the previous iteration. After the first iteration of the method over $k_o = 1$ to k, this approach requires less computation than determining the $\lceil n_o / k_o \rceil$ points closest to i. Because of the increased speed of using a definition of Proximity(i) based on U instead of $\lceil n_o / k_o \rceil$, a value of U may be selected earlier, after examining only one or a few $k_o$ values, and subsequently adjusted at each new iteration according to the numbers of elements admitted to Proximity(i) on the preceding iteration. Referring to U may exhaust all elements of N before $H_o$ contains k elements, and if U is estimated to be of reasonable size, this may signal that k is selected too large.

## B. Intensity-Based Starting Methods

## 1. The Primary MinSum Starting Method

We now consider the intensity-based starting method associated with the MinSum-centroid. In the following we omit explanatory comments for pseudocode that is made clear by the observations for the corresponding Primary MinMax Starting Method.

**Primary MinSum Method**

*Initialization*: Begin with the ordered indexes i(1), i(2), …, i(n-1) for each i ∈ N and set $N_o = N$, $n_o = n$ and $\delta(j) = True$, j ∈ N.

For $k_o = k$ to 1
    ClusterSize $= \lceil n_o / k_o \rceil$
    ScanSize = ClusterSize − 1
    Execute the **MinSum Distance Algorithm**
        (At the conclusion: i* = arg min(Max (d(i,,j): j ∈ C) (= MSSpan(C)) over all sets
        C ⊂ $N_o$ such that |C| = ClusterSize.)
    **Generate C($k_o$) and update the set $N_o$**
    (This portion of the Primary MinSum Method is identical to that of the Primary MinMax
    Method, except that MSCentroid($k_o$) = i* replaces MMCentroid($k_o$) = i*.)
*Endfor*

The internal MinSum Distance algorithm is also nearly a replica of the internal MinMax Distance algorithm upon replacing BestDistance with a corresponding value BestSum. All arrays and values of the MinSum Distance Algorithm are shared with the Primary MinSum Starting Method.

**MinSum Distance Algorithm**
*Conventions:*
BestSum = the MinSum value $\sum(d(i,j): \; = d(i^*,j)$ associated with the best candidate for i*.

*Initialization:*
BestSum = *Large*

For i ∈ $N_o$:
    Sum = 0
    Scan = 0
    *For* s = 1 *to* n − 1
        j = i(s)
        *If* ($\delta(j)$ = *True*) *then*
            Sum := Sum + d(i,j)
            Scan := Scan + 1
            *If* (Scan = ScanSize) *Break*
        *Endif*
    *Endfor*
    *If* (Sum < BestSum) *then*
        i* = i
        BestSum = Sum
    *Endif*

*Endfor*

An analog of Theorem 1 applies to this Primary MinSum Starting Method, which we do not bother to state here.

## 2. Advanced Intensity-Based Starting: The Adaptive MinSum Starting Method

The description of the "external" portion of the Adaptive MinSum method, given next, is nearly identical to that of the Adaptive MinMax Starting Method. However, the internal Adaptive Phase 1 and Phase 2 algorithms differ more significantly from the MinMax case, as indicated below.

**Adaptive MinSum Method**
*Initialization*: Begin with the ordered indexes i(1), i(2), …, i(n-1) for each $i \in N$ and set $N_o = N$, $n_o = n$ and $\delta(j) = \textit{True}$, $j \in N$.

For $k_o = k$ to 1
    MinSize $= \lceil n_o / k_o \rceil$
    MaxSize $= n_o - \text{MinSize} \cdot (k_o - 1)$
    MinScan = MinSize $- 1$
    MaxScan = MaxSize $- 1$
    Execute the **Phase 1 MinSum Algorithm** (to identify BestSum and associated values)
    Execute the **Phase 2 MinSum Algorithm** (to identify i*)
    **Generate $C(k_o)$ and update the set $N_o$**
    (This portion of the Adaptive MinSum Method is identical to that of the Primary
    MinSum Method, except that "*If* (Scan = BestScan) *Break*" replaces
    "*If* (Scan = ScanSize) *Break*".)
*Endfor*

**Phase 1 MinSum Algorithm**
*Assumption:* $N_o$ contains at least 2 elements
BestDistance $= \textit{Large}$
BestMinGap $= 0$
BestSum $= \textit{Large}$

For $i \in N_o$
    MinGap $= \textit{Large}$
    Sum $= 0$
    SumGap $= 0$
    Scan $= 0$
    *For* s $= 1$ *to* n $- 1$
        j = i(s)

        *If ($\delta(j)$ = True) then*

            Scan := Scan + 1

            Distance = d(i,j)

            Sum := Sum + Distance

            *If (Scan > 1) then*

                Gap = Distance − PreviousDistance

                SumGap = SumGap + Gap

                MinGap = Min(Gap, MinGap)

            *Endif*

            *If (Scan = MinScan) then*

                Sum(i) = Sum (initialize Phase 2 at Sum = Sum(i))

                s(i) = s (initialize Phase 2 at s = s(i))

                *Break*

            *Endif*

            PreviousDistance = Distance

        *Endif*

    *Endfor*

    *If ((Sum < BestSum)*

        *or ((Sum = BestSum) and (Distance < BestDistance))*

        *or ((Sum = BestSum) and (Distance = BestDistance)*

            *and (MinGap > BestMinGap)) ) then*

        BestSum = Sum

        BestDistance = Distance

        BestSumGap = SumGap

        BestMinGap = MinGap

    *Endif*

Endfor

BestMeanGap = BestSumGap/(MinScan − 1)

TargetGap = $\lambda$·BestMeanGap + (1 − $\lambda$)·BestMinGap) (suggested default, $\lambda$ = .3)

*(The following 3 values are inherited to apply for each i $\in$ N$_o$ in Phase 2)*

DistanceLimit = BestDistance + TargetGap (applies to MinMax version)

FirstSumLimit = BestSum + TargetGap

DeltaSum = DistanceLimit

**Phase 2 Adaptive MinSum Algorithm**

*Conventions:*

Sum = Sum d(i,j) using the ordered scanning of elements from i $\in$ N$_o$

BestSum = the Sum value for the current best candidate for i*.

DistanceLimit = the largest distance allowed for d(i,j) when scanning from element i.

SumLimit = the limit on Sum, increasing by DeltaSum at each Scan > MinScan

BestScan = the value of Scan that produces the best candidate for i*.

*Inherited from Phase 1:*

DistanceLimit, FirstSumLimit, DeltaSum and s(i), Sum(i), $i \in N_o$

*Initialization:*
BestDistance = *Large*
BestScan = *Large*
BestSum = *Large*

For $i \in N_o$
    Scan = MinScan − 1 (Scan will be immediately incremented to MinScan)
    SumLimit = FirstSumLimit
    Sum = Sum(i)
    *For* s = s(i) *to* n − 1
        j = i(s)
        *If* ($\delta$(j) = *True*) *then* (automatically true for s = s(i))
            Scan := Scan + 1 (reaches MinScan on first execution when s = s(i))
            *If* (Scan > MinScan) *then*
                Sum := Sum + d(i,j)
                SumLimit := SumLimit + DeltaSum
            *Endif*
            *If* (Sum > SumLimit) *then*
                Scan := Scan − 1
                *Break* (exit the "For loop")
            *Endif*
            *If* (Scan = MaxScan) *Break*
        *Endif*
    *Endfor*
    *If* (Scan ≥ MinScan) *then* (i is a candidate for i*)
        *If* ((Scan > BestScan)
            *or* ((Scan = BestScan) *and* (Sum < BestSum))) *then*
            i* = i
            BestSum = Sum
            BestScan = Scan
        *Endif*
    *Endif*
*Endfor*

The analog of Theorem 2 and its proof that applies to the Adaptive MinSum Starting Method gives BestSum the role taken by BestDisance, but otherwise establishes the same double layer of priorities to assure the final $C_{best}(v^*)$ (for $v^* = $ BestScan + 1) and its associated point i* give the cluster and MinSum-centroid we seek.